# Sequential & Parallel Algorithms for Big-Integer Numbers Subtraction

Youssef Bassil, Aziz Barbar

*GJCST Classifications:*
*C.1.4, C.1.2, F.1.3*

*Abstract-* Many emerging computer applications require the processing of large numbers, larger than what a CPU can handle. In fact, the top of the line PCs can only manipulate numbers not longer than 32 bits or 64 bits. This is due to the size of the registers and the data-path inside the CPU. As a result, performing arithmetic operations such as subtraction on big-integer numbers is to some extend limited. Different algorithms were designed in an attempt to solve this problem; they all operate on big-integer numbers by first converting them into a binary representation then performing bitwise operations on single bits. Such algorithms are of complexity O(n) where n is the total number of bits in each operand.

This paper proposes two new algorithms for performing arithmetic subtraction on big-integer numbers. The two algorithms are different in that one is sequential while the other is parallel. The similarity between them is that both follow the same concept of dividing the big-integer inputs into several blocks or tokens of 60 bits (18 digits) each; thus reducing the input size n in O(n) by a factor of 60. Subtraction of corresponding tokens, one from each operand, is performed as humans perform subtraction, using a pencil and a paper in the decimal system.

Both algorithms are to be implemented using MS C#.NET 2005 and tested over a multiple processor system. Further studies can be done on other arithmetic operations such as addition and multiplication.

## I  Introduction

Contemporary PCs usually handle and operate on numbers not longer than 32 bits and 64 bits (Maxfield & Brown, 2004). The reason behind this is that PCs' CPUs can only accommodate and manipulate numbers of that length (Hennessy & Patterson, 2006). The real problem arises when certain applications require performing computer arithmetic such as addition, subtraction, multiplication, and division on numbers larger than 64 bits, at high speed. For instance, in cryptography cipher keys can be as large as 512-bits and 1024-bits. In banking systems, customer's balances can be sometimes larger than 64-bits taking into consideration the difference between currencies.

Manuscript received "28th December"
Youssef Bassil, Department of Computer Science, American University of Science & Technology, Beirut, Lebanon,
Aziz Barbar, Department of Computer Science, American University of Science & Technology, Beirut, Lebanon,
Telephone: 961 1 218716 ext 311

In some scientific and mathematical applications, performing precise and accurate real-time computations demand the use of numbers larger than 64 bits.

Various solutions were proposed to solve this problem; the majority of them carry out arithmetic operations on the bit level using bitwise operations (Knuth, 1997), (Koren, 2001). Furthermore, none of them is designed to exploit multiprocessor systems and shared memory architecture. The complexity of such algorithms is usually O(n) where the basic operation is executed n times, that is eventually equal to the number of bits in the big-integer input.

In this paper, we are proposing a sequential and a parallel algorithm for handling arithmetic subtraction on big-integer numbers. Both algorithms carry out subtraction on 60 bits unit tokens, and not on individual bits as other existing approaches. Accordingly, the algorithm is of best-case complexity O(n) with n reduced by a factor of 60; hence, the computation is 60 times faster. In general, both algorithms emulate the elementary pen and paper method used to perform subtraction in the decimal system in that they start by dividing the big-integer operands into tokens or blocks of 60 bits each. Then each two corresponding tokens are subtracted from each other while the borrows are handled properly. In the sequential algorithm, this whole process is executed on a single processor system; while in the parallel algorithm, each two corresponding tokens are assigned to a particular processor in a multi-processor system to be subtracted from each other. Experiments showed a momentous improvement over other existing techniques and approaches.

## II  Existing solutions

Many programming libraries were developed to solve the problem of performing arithmetic calculations over big-integer numbers. Some of them are proprietary third party dynamic link libraries (DLL), either available for free or sold at a given cost; while others are shipped as a part of the programming language application programming interface (API). For instance, the MS .NET Framework 4.0 provides the BigInteger class in the namespace System.Numerics (MSDN, 2009). The Java programming language provides another BigInteger class in the java.math package (Java Documentation, 2008). Both carry out arithmetic operations on big-integer numbers using bitwise operations (Java BigInteger Source-code, 2006). They first convert the base-10 big-integer input to a base-2 binary representation, then they employ the bitwise operators OR and XOR to perform binary subtraction over string of bits.

The algorithm behind these libraries is of complexity O(n) where n is the total number of bits constituting each operand. In terms of time efficiency, the number of times the basic operation is executed, is equal to the number of bits in the big-integer operands. Moreover, most of these libraries are not designed to work in a parallel fashion; they are only optimized to operate over single-processor systems.

### III THE PROPOSED SEQUENTIAL ALGORITHM

The sequential algorithm proposed in this paper is based on the same principle humans use to perform subtraction, using a pencil and a paper in the decimal system. Generally speaking, inputs of big-integer numbers are chopped to several smaller tokens each made out of 60 bits (18 digits). Afterwards, each two corresponding tokens are treated as single units and aligned on top of each others. Then they are subtracted from each others while handling correctly the borrows. It is worth noting that no conversion to base-2 is to occur, the computation is totally done in the base-10 decimal system

Below is a list of steps executed by the sequential algorithm to subtract two big-integer numbers:

i. Two big-integer operands of type string a and b, such as a is greater or equal to b, are fed to the algorithm. SubtractBigInteger(a , b)
ii. Both string operands a and b are then parsed and divided from right to left into smaller chunks or tokens $t_i(p)$ where i is the token index and p is the operand to which ti belongs. Consequently, operand $a = t_{n-1}(a)… t_0(a)$ and operand $b = t_{m-1}(b)… t_0(b)$, where n and m are the total number of tokens constituting each of the operands. The length of each single produced token ti is less than or equal to 18. (In the C# programming language, the largest integer data type is long (signed by default) which can store up to 19 digits or $2^{63}=$ 9223372036854775808. Since in mathematical subtraction there exist the concept of a borrow, it is crucial to reserve 1 digit for a possible borrow, resulting in 19-1=18 digits represented by 60 bits). The resulting tokens will be stored as strings in two arrays, each for a particular operand.
iii. The tokens contained in the two arrays are to be converted from *string* to *long* data type. In other words, each single token, now representing an array element with a maximum length of 18 digits, is to be converted to an integer value of type *long*. The conversion is required because arithmetic subtraction cannot be performed over *string* types
iv. Both arrays, now containing long type tokens, are aligned on top of each other. Starting from the rightmost token, each two corresponding tokens are subtracted from each other as in performing subtraction using a pencil and a paper: $t_i(c) = t_i(a) - t_i(b)$. If $t_i(a) < t_i(b)$, then a borrow of 1 must be subtracted from $t_{i+1}(a)$. $t_{i+1}$ is the next token on the left of the two tokens being currently subtracted. Consequently, $t_{i+1}(a)$ would be equal to $t_{i+1}(a)-1$ and a 1 representing the borrow is appended as the 19th digit to $t_i(a)$. It is important to mention that in case $t_{i+1}(a)$ is equal to 0, a borrow is taken from $t_{i+2}(a)$ then propagated to $t_i+1(a)$ and then to $t_i(a)$. Under special cases, a borrow can be propagated from $t_{n-1}(a)$ to ti(a). Since operand a is always greater or equal to operand b, tn-1(a) must be able to provide a borrow in a way or another. Now $t_i(a) >= t_i(b)$ and $t_i(a) - t_i(b)$ is feasible to be calculated.
v. Finally, all the produced ti(c) are to be concatenated together to attain result = $t_{r-1}(c)… t_0(c)$. It is important to note that this algorithm can handle operands of different sizes, in a sense that excessive tokens, which should logically belong to the largest operand are just appended to the final result. Figure 1 summaries the different steps performed by the sequential algorithm in order to subtract two operands a and b.

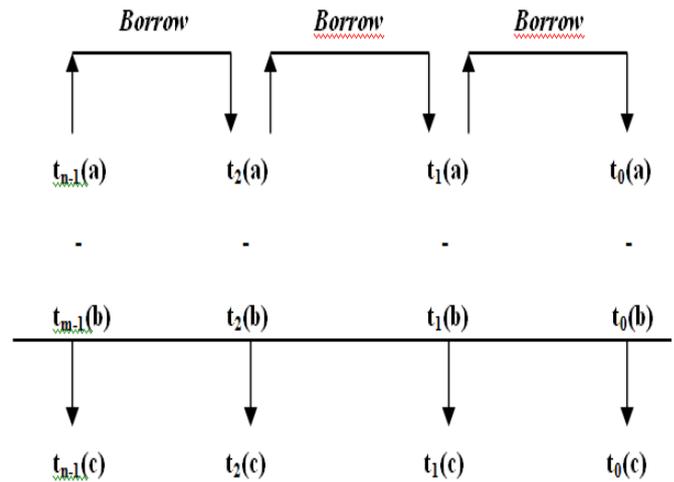

Fig.1. Subtracting two operands using the proposed sequential algorithm.

#### A. Implementation

Below is the code of the sequential algorithm implemented using MS C#.NET 2005 under the .NET Framework 2.0 and MS Visual Studio 2005.

```
private string SubtractBigInteger(string a, string b)
{
    long[] tokens_A = ParseOperand(a);
    long[] tokens_B = ParseOperand(b);

    int length = tokens_A.Length ;

    long[] result = new long[length];

    int i, j;
    for (i = length - 1, j = length - 1; j != -1; i--, j--)
    {
        // we must borrow a 1 from the token on the left
        while (tokens_A[i] < tokens_B[j])
        {
```

```
            int k = i - 1;

         while (true)
          {
            if (tokens_A[k] != 0)
            {
             tokens_A[k]--;
              // Adding the 19th digit to the left of the
                        token
              // that needs borrow
              tokens_A[k + 1] = tokens_A[k + 1] + (1 *
                        1000000000000000000);
             break;
            }
            else k--;
          }
         }

          // Performing the subtraction
          result[i] = tokens_A[i] - tokens_B[j];
        }

         return ConvertToString(result);
       }

private long[] ParseOperand(string operand)
   {
       ArrayList list = new ArrayList();

       for (int i = 0; operand.Length > 18; i++)
         {
           list.Add(operand.Substring(operand.Length - 18));

       operand = operand.Substring(0, operand.Length -18))
}
       list.Add(operand);
        list.Reverse();

        long[] tokens = new long[list.Count];

        for (int j = 0; j < tokens.Length; j++)
         {
            tokens[j] = Convert.ToInt64(list[j]);
         }
          return tokens;
        }
```

### B.  *Experiments and Results*

We will be comparing in our testing the execution time of the proposed sequential algorithm with the System.Numercis.BigInteger class included in MS .NET Framework 4.0, and the java.math.BigInteger class included in Java SE 1.6.

Below are two code segments that illustrate how to use the methods of the built-in classes System.Numercis.BigInteger and java.math.BigInteger in order to subtract two big-integer numbers using the C#.NET and the Java language.

```
using System.Numerics;
public class BigIntegerTest_Csharp
  {
     public static void Main(string args[])
{
     String operandA =
"12345678909876543211234567890987654321" ;
     String operandA =
"12345678909876543211234567890987654321" ;
      BigInteger a = BigInteger.Parse(operandA) ;
      BigInteger b = BigInteger.Parse(operandB) ;

      BigInteger results = BigInteger.Subtract(a, b);
      Console.WriteLine(results.ToString());
    }
}

import java.math.BigInteger;
 public class BigIntegerTest_Java
 {
    public static void main(String args[])
    {

       String operandA =
"12345678909876543211234567890987654321" ;
       String operandA =
"12345678909876543211234567890987654321" ;
      BigInteger a = new BigInteger(operandA) ;
      BigInteger b = new BigInteger(operandB) ;

      System.out.print("" + a.subtract(b)) ;
   }
}
```

As a testing platform, we are using a desktop IBM-compatible PC with Intel Core single core processor with 1.66 MHz clock speed, 256KB of cache, and 512MB of RAM. The operating system used is MS Windows XP Professional SP2.

It is worth noting that the execution time obtained for all different algorithms is an average time obtained after five consecutive runs of the same test.

| Test Case | Operands | Value | Value Length |
|---|---|---|---|
| 1 | A | X | 20,000 base-10 digits |
| 1 | B | Y | 20,000 base-10 digits |
| 2 | A | X | 100,000 base-10 digits |
| 2 | B | Y | 100,000 base-10 digits |
| 3 | A | X | 500,000 base-10 digits |
| 3 | B | Y | 500,000 base-10 digits |
| 4 | A | X | 1000,000 base-10 digits |
| 4 | B | Y | 1000,000 base-10 digits |

Table 1: Test cases.

| Test Case | Operation | Results | Execution Time in Seconds |
|---|---|---|---|
| 1 | A-B | X-Y | 2.01 |
| 2 | A-B | X-Y | 17.19 |
| 3 | A-B | X-Y | 514.33 |
| 4 | A-B | X-Y | 2487.05 |

Table 2: Results obtained for the .NET class.

| Test Case | Operation | Results | Execution Time in Seconds |
|---|---|---|---|
| 1 | A-B | X-Y | 0.51 |
| 2 | A-B | X-Y | 8.02 |
| 3 | A-B | X-Y | 165.34 |
| 4 | A-B | X-Y | 601.90 |

Table 3: Results obtained for the Java class.

| Test Case | Operation | Results | Execution Time in Seconds |
|---|---|---|---|
| 1 | A-B | X-Y | 0.18 |
| 2 | A-B | X-Y | 2.06 |
| 3 | A-B | X-Y | 66.04 |
| 4 | A-B | X-Y | 318.71 |

Table 4: Results obtained for our sequential algorithm.

From the obtained results, delineated in tables 1-4, it became clear that our sequential algorithm outsmarted all other algorithms in all different test cases. When big-integer numbers were in length respectively 20,000 and 100,000 digits, our algorithm beat the .NET and Java classes with few little seconds. However, when numbers became as large as 500,000 digits, our algorithm surpassed the Java class by around 100 seconds (1.6 minutes) and the .NET class by around 450 seconds (7.5 minutes). Furthermore, our proposed algorithm increased the pace between its rivals when the length of operands reached the 1,000,000 digits. It surpassed the Java class by around 283 seconds (4.7 minutes), and the .NET class by around 2169 seconds (36.1 minutes).

### C. Algorithm Analysis

The sequential algorithm showed a real speed improvement over other existing approaches. It outperformed the .NET and Java built-in classes by several seconds and this gap exponentially increased as the length of the big-integer operands became larger. This speed improvement is due to the reduction of the input size $n$ in $O(n)$. The .NET, Java, and our proposed algorithm are all of best-case complexity $O(n)$. However, the $n$ in the .NET and Java algorithms represents the total number of bits in each operand; while in our proposed algorithm, $n$ represents the total number of tokens in each operand. For instance, the decimal number 999999999999999999 (18 digits) is represented in base-2 as 110111100000101101101011001110100111011000111111 111111111111 (60 bits). This makes $n=60$ and thus the basic operation is executed 60 times. On the other hand, in our proposed algorithm, this whole 999999999999999999 is treated as a single unit token. This makes $n=1$ and thus the basic operation is executed only 1 time. As a result, the time efficiency of our algorithm is supposedly to be 60 times faster than the other algorithms. However, this is not the case, since handling the borrows requires various extra operations to be executed; a fact that imposes further processing overhead, and increases the computation time. Accordingly, the best-case is when no borrows are needed throughout the execution of the whole algorithm, then the basic operation is executed $n$ times where $n$ is the total number of tokens and this makes $C_{Best}(n)=n$ belonging to $O(n)$.

### IV THE PROPOSED PARALLEL ALGORITHM

The parallel algorithm proposed in this paper is a multithreaded parallel algorithm designed to be executed over multi-processor shared memory architecture. It is based on the principle of performing arithmetic subtraction as humans perform subtraction, using a pencil and a paper in the decimal system. Ordinarily, the algorithm starts by breaking down big-integer numbers into blocks or tokens of 60 bits each. Then subtraction starts in a sequence of multiple iterations. On the first iteration, each two corresponding tokens are assigned to a particular thread, which subtracts them from each others using a particular microprocessor. When a borrow is needed, the algorithm assumes that the borrow is there. For instance, if token 99 is to be subtracted from token 88, the algorithm will directly subtract 99 from 188 as if a borrow had occurred. Then, a value of 1 representing the borrow is stored in a shared array. On the second iteration, that borrow will be subtracted accordingly from the token on the left of the previous result. Iterations continue until no more borrows are generated from a previous iteration.

Below is a list of steps executed by the parallel algorithm to subtract two big-integer numbers:

i. Two very large string numbers operand a and operand b, such as a is greater or equal to b, are fed to the algorithm. SubtractBigInteger_Parallel(a , b)

ii. Both *string* operands a and b are then parsed and divided from right to left into smaller chunks or tokens $t_i(p)$ where i is the token index and p is the operand to which $t_i$ belongs. Consequently, operand a = $t_{n-1}(a)$… $t_0(a)$ and operand b = $t_{m-1}(b)$… $t_0(b)$ where n and m are the total number of tokens constituting each of the operands. The length of each single produced token $t_i$ is less than or equal to 18. (In the C# programming language, the largest integer data type is *long* (signed by default) which can store up to 19 digits or $2^{63}$= 9223372036854775808. Since in mathematical subtraction there exist the concept of a borrow, it is crucial to reserve 1 digit for a possible *borrow*, resulting in 19-1=18 digits represented by 60 bits). The resulting tokens will be stored as *string* in two arrays, each for a particular operand.

iii. The tokens contained in the two arrays are to be converted from string to long data type. In other

words, each single token, now representing an array element with a maximum length of 18 digits, is to be converted to an integer value of type long. The conversion is required because arithmetic subtraction cannot be performed on string types

iv. Each processor pi in a multiprocessor system is assigned two tokens, one from each operand. Therefore the processor pi is assigned tokens $t_i(a)$ and $t_i(b)$ with the purpose of calculating $t_i(c) = t_i(a) - t_i(b)$. For instance, p0 will calculate $t_0(c)$, p1 will calculate $t_1(c)$, p2 will calculate $t_2(c)$ and so on. We are to assume that the number of processor is equal to the number of tokens; otherwise, tokens are distributed equally among processors. For instance, if the number of processors is half the number of tokens, each processor will be assigned 4 tokens (2 from each operand) to be calculated as in sequential approach. $t_i(c) = t_i(a) - t_i(b)$ and then $t_{i+1}(c) = t_{i+1}(a) - t_{i+1}(b)$

v. A borrow is handled using multiple processing iterations and a shared array called borrows[0...n-1] used to store all the produced borrows. For that reason, we have added a new variable called T as in $t_i(c,T)$ to represent the iteration into which $t_i(c)$ is being calculated. T=1 is the first iteration and T=n is the nth iteration. In this approach, in case a borrow was needed, the algorithm assumes that the borrow is there. For instance, if token 99 is to be subtracted from token 88, the algorithm will directly subtract 99 from 188 as if a borrow had occurred, and borrows[i+1] is set to 1. It is i+1 so that on the next iteration T=2, borrows[i+1] will be correctly subtracted from the previously calculated $t_{i+1}(c,1)$. Likewise, if another borrow is needed for $t_i(c,2)$, borrows[i+1] is set to 1 overwriting any previous value. Consequently, on the next iteration (T=3) borrows[i+1] will be correctly subtracted from $t_{i+1}(c,2)$. This will keep on looping until no more borrows are generated (borrows[0...n-1] contains no 1's). As an example, if on the first iteration (T=1), a borrow was needed for t4(a,1), then a borrow is assumed to have occurred, t4(c,1) is calculated and borrows[5] is set to 1, p5 (processor 5) starts a second iteration (T=2) in an attempt to calculate $t_5(c,2) = t_5(c,1) - borrows[5]$. In the meantime, all other pi where borrows[i]=0 will refrain from executing. If after T=2 no more borrows are needed, the loop process stops.

vi. Finally, all the $t_i(c)$ produced after many iterations are to be concatenated together: result = $t_{n-1}(c)…t_0(c)$. Figure 2 summaries the different steps performed by the parallel algorithm in order to subtract two operands a and b.

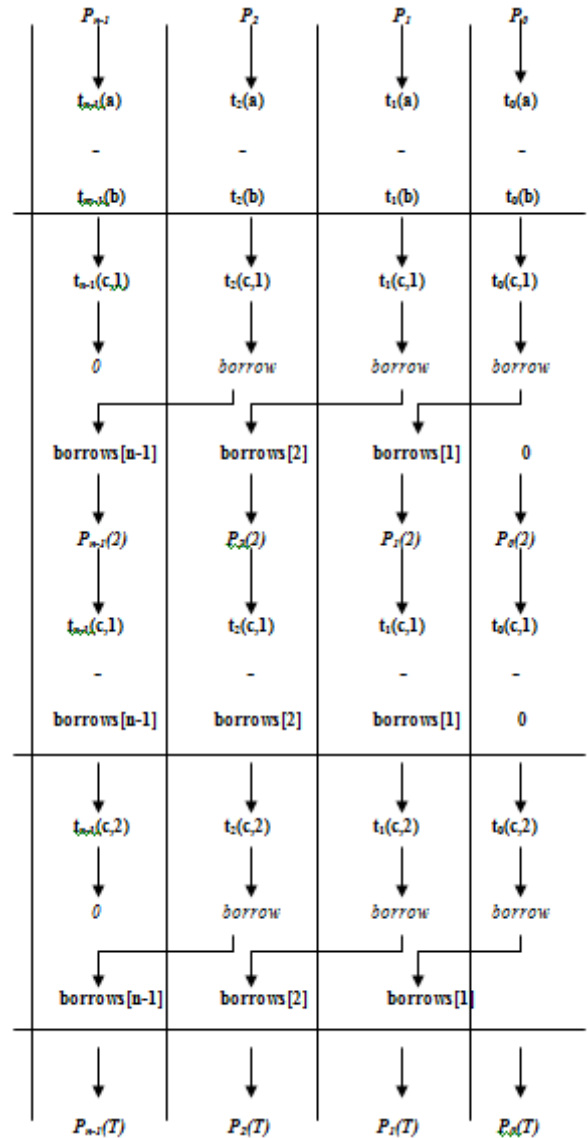

Fig. 2. Subtracting two operands using the proposed parallel algorithm.

A. Implementation

Below is the code of the proposed parallel algorithm implemented in MS C#.NET 2005 under the .NET Framework 2.0 and MS Visual Studio 2005. It uses classes and methods from *System.Threading* namespace to create, destroy and execute threads. All threads can read and write to a shared memory space where tokens, carries, flags and other variables are stored and shared.

```
long[] tokens_A ;
long[] tokens_B ;

long[] result;
int[] borrows;

int numberOfProcessors;

int sharedIndex;
```

```csharp
    int terminatedThreads=0;
    int T=1;
    Thread[] threads;

  public void SubtractBigInteger_Parallel(string a,
          stringb)
  {
    tokens_A = ParseOperand(a, 18);
    tokens_B = ParseOperand(b, 18);

    result = new long[tokens_A.Length];

    // By default borrows is populated with 0s
    borrows = new int[tokens_A.Length];

    numberOfProcessors = GetNumOfProcessors();

    threads = new Thread[numberOfProcessors];

    CreateThreads();
  }

  private void CreateThreads()
  {
    sharedIndex = numberOfProcessors;

    for (int i = 0; i < numberOfProcessors; i++)
    {
      threads[i] = new Thread(new
       ThreadStart(Process));
      threads[i].Start();
    }
  }

  private void Process()
  {
    int index = sharedIndex--;

    if(T==1) // First iteration
    {
        if(tokens_A[index] < tokens_B[index])
        {
          // Add a borrow
          tokens_A[index] + (1
            * 1000000000000000000);

          borrows[index-1] = 1 ;
      }

      else borrows[index-1] = 0 ;

    result[index] = tokens_A[index] - tokens_B[index];
  }
   else
{
    if(result[index] == 0) // result[index] < borrows[index]
    {
      // Add a borrow
        result[index] + (1 * 1000000000000000000);

      borrows[index-1] = 1 ;
    }
    else borrows[index-1] = 0 ;

    result[index] = result[index] - borrows[index];
}

    terminatedThreads++;

    IsProcessingDone();
  }

  private void IsProcessingDone()
  {
    if (terminatedThreads == numberOfProcessors)
    {
      if (AreMoreBorrows())
      {
        T++ ;

        // Creates new set of threads in the next iteration
        CreateThreads();
      }
      else DisplayResults();
    }
  }

  private bool AreMoreBorrows()
  {
    for (int i = 0; i < borrows.Length; i++)
    {
      if (borrows[i] = = 1)
        return true;
    }

    return false;
  }

  private string DisplayResults()
  {
    return ConvertToString(result);
  }

  private long[] ParseOperand(string operand)
  {
    ArrayList list = new ArrayList();

    for (int i = 0; operand.Length > 18; i++)
    {
      list.Add(operand.Substring(operand.Length - 18));
   operand = operand.Substring(0, operand.Length-    18);
}
    list.Add(operand);
```

```
    list.Reverse();

    long[] tokens = new long[list.Count];

    for (int j = 0; j < tokens.Length; j++)
    {
        tokens[j] = Convert.ToInt64(list[j]);
    }

    return tokens;
}
```

### B. Experiments and Results

In this section, a comparison of the execution time between the sequential and the parallel algorithm, both proposed in this paper, is undertaken using a desktop IBM-compatible PC with 4 processors of type Intel Core single core with 1.8 MHz clock speed, 512KB of cache, and 2GB of RAM. The operating system used is MS Windows Server 2003 SP1.

It is important to note here that the execution time obtained for all different algorithms is an average time obtained after five consecutive runs of the same test.

| Test Case | Operands | Value | Value Length |
|---|---|---|---|
| 1 | A | X | 20,000 base-10 digits |
| 1 | B | Y | 20,000 base-10 digits |
| 2 | A | X | 100,000 base-10 digits |
| 2 | B | Y | 100,000 base-10 digits |
| 3 | A | X | 500,000 base-10 digits |
| 3 | B | Y | 500,000 base-10 digits |
| 4 | A | X | 1000,000 base-10 digits |
| 4 | B | Y | 1000,000 base-10 digits |

Table 5: Test cases.

| Test Case | Operation | Results | Execution Time in Seconds |
|---|---|---|---|
| 1 | A+B | X+Y | 0.10 |
| 2 | A+B | X+Y | 2.12 |
| 3 | A+B | X+Y | 64.51 |
| 4 | A+B | X+Y | 311.66 |

Table 6: Results obtained for our sequential algorithm.

| Test Case | Operation | Results | Execution Time in Seconds |
|---|---|---|---|
| 1 | A+B | X+Y | 0.03 |
| 2 | A+B | X+Y | 0.92 |
| 3 | A+B | X+Y | 23.17 |
| 4 | A+B | X+Y | 91.96 |

Table 7: Results obtained for our parallel algorithm.

The results delineated in tables 5-7 show that the parallel algorithm outperformed the sequential algorithm by an average factor of 3.2. At the beginning, when operands were in length 20,000 and 100,000 respectively, the difference was not that evident. However, when numbers became larger, the gap increased and the execution time was speeded up by around 320%.

### C. Algorithm Analysis

The parallel algorithm improved on the sequential algorithm and boosted its execution time by around 320%. In terms of algorithm complexity, and assuming that every token is exactly assigned to a particular processor, the best-case efficiency is when no borrows are generated after the first iteration; a fact that achieves the best performance, and thus $C_{Best}(n)=1$, that is each processor executes the basic operation only one time. The worst-case efficiency is when a new borrow is generated after each iteration, this would require n-1 iterations in order to propagate and subtract all the borrows and thus $C_{Worst}(n)=n-1$. Consequently, The average-case efficiency is $C_{Average}(n)=(n-1)/2$

## V  FUTURE WORK

Future research can improve upon our proposed algorithms so much so that other arithmetic operations such as addition, multiplication and division are added. Besides, a distributed version of the same algorithms is to be designed so that it can be executed over a network of regular machines, making the implementation less expensive and more scalable.

## VI  ACKNOWLEDGMENTS